\PassOptionsToPackage{table}{xcolor}
\documentclass[sigconf, nonacm, pdfa]{acmart}

\newcommand\vldbdoi{10.14778/3827998.3828016}
\newcommand\vldbpages{4063 - 4075}
\newcommand\vldbvolume{19}
\newcommand\vldbissue{12}
\newcommand\vldbyear{2026}

\newcommand\vldbauthors{{Yifan Wu, Yuhan Li, Zhenhua Wang, Ke Chen, Lidan Shou, Zonghao Chen, Liang Lin, Huan Li, Gang Chen}}
\newcommand\vldbtitle{ScaleSense: Cost-Intelligent Scaling Framework via Learned Resource Estimation in Alibaba AnalyticDB}
\newcommand\vldbavailabilityurl{}
\newcommand\vldbpagestyle{empty}

\usepackage{amsmath}
\usepackage{graphicx}
\usepackage{textcomp}
\usepackage{xcolor}
\usepackage[normalem]{ulem}
\usepackage{listings}
\usepackage[linesnumbered,boxed,ruled,vlined,commentsnumbered]{algorithm2e}
\usepackage{courier}
\usepackage{booktabs}
\usepackage{color}
\usepackage{multirow}
\usepackage{makecell}
\usepackage{wrapfig}
\usepackage{url}
\usepackage{ifthen}
\usepackage{balance}
\usepackage{forest}
\usepackage{pifont}
\usepackage{adjustbox}
\usepackage[shortlabels]{enumitem}
\usepackage{soul}
\usepackage{xcolor}
\usepackage{multicol}
\usepackage{xparse}
\usepackage{mdframed}
\usepackage{bookmark}
\usepackage[utf8]{inputenc}

\usepackage{arydshln}
\usepackage{amsthm}

\SetKwProg{Fn}{Function}{}{end}
\SetKwProg{Proc}{Procedure}{}{end}

\definecolor{mymauve}{rgb}{0.58,0,0.82}
\definecolor{dkgreen}{rgb}{0,0.6,0}
\definecolor{browncolor}{rgb}{0.6, 0.3, 0.0}

\definecolor{lowdpcolor}{rgb}{0.9, 0.8, 0.2}
\definecolor{lightpurple}{rgb}{0.8, 0.7, 0.9}
\definecolor{lightpink}{rgb}{1.0, 0.8, 0.86}

\definecolor{advantagecolor}{rgb}{0.8, 1.0, 0.8}%
\definecolor{disadvantagecolor}{HTML}{FFE5E5}%

\definecolor{lowdpcolor}{HTML}{DDDDFF}
\definecolor{highdpcolor}{HTML}{FFCCCC}

\definecolor{lightred}{HTML}{FFCCCC}%
\definecolor{lightblue}{HTML}{CCCCFF}%

\setenumerate[1]{itemsep=0pt,partopsep=0pt,parsep=\parskip,topsep=0.5pt}
\setitemize[1]{itemsep=0pt,partopsep=0pt,parsep=\parskip,topsep=0.5pt}
\setdescription{itemsep=0pt,partopsep=0pt,parsep=\parskip,topsep=0.5pt}

\definecolor{myblue}{rgb}{0,0,1}
\definecolor{verylightblue}{rgb}{0.8, 0.9, 1.0}
\sethlcolor{verylightblue}

\forestset{
  default preamble={
    for tree={
      parent anchor=south,
      child anchor=north,
      align=center,
      edge={-latex},
      rounded corners,
      draw,
      fill=white,
      s sep=7mm,
      l sep=10mm,
      anchor=center,
      calign=center,
      align=center
    }
  }
}

\usepackage{caption}
\def\BibTeX{{\rm B\kern-.05em{\sc i\kern-.025em b}\kern-.08em
    T\kern-.1667em\lower.7ex\hbox{E}\kern-.125emX}}

\newtheorem*{scenario*}{\bf Targeted Scenarios}

\newtheorem{problem}{\bf Problem}

\newtheorem{insight}{\bf Insight}

\newcommand{\ScaleSense}{\textsc{ScaleSense}\xspace}

\useunder{\uline}{\ul}{}

\usepackage[skins,breakable]{tcolorbox}

\renewcommand{\shortauthors}{Yifan Wu et al.}

\usepackage[utf8]{inputenc}
\usepackage[UKenglish]{babel}
\usepackage{colorprofiles}
\usepackage[a-2b]{pdfx}
\usepackage{fontawesome}
\usepackage[T1]{fontenc}
\usepackage{pifont}

\hypersetup{
  pdfstartview=,
  colorlinks=false,
  pdfborder={0 0 0},
  hidelinks,
  pdftitle={ScaleSense: Cost-Intelligent Scaling Framework via Learned Resource Estimation in Alibaba AnalyticDB},
  pdfauthor={Yifan Wu},
  pdfsubject={ScaleSense: Cost-Intelligent Scaling Framework via Learned Resource Estimation in Alibaba AnalyticDB},
  pdfkeywords={Cloud Data Warehouse, Database, AI, Serverless}
}

\begin{document}

\title{\ScaleSense: Cost-Intelligent Scaling Framework via Learned Resource Estimation in Alibaba AnalyticDB}

\settopmatter{authorsperrow=3}
\author{Yifan Wu}
\affiliation{
\institution{Zhejiang University$^\dag$}
}
\email{yifan.wu@zju.edu.cn}

\author{Yuhan Li}
\author{Zhenhua Wang}
\affiliation{
\institution{Alibaba Cloud Computing}
}
\email{{lyh200442, wzh420090}@alibaba-inc.com}

\author{Ke Chen}
\author{Lidan Shou}
\authornotemark[1]
\author{Zonghao Chen}
\affiliation{
\institution{Zhejiang University$^\dag$}
}
\email{{ck, should, zhchen.cs}@zju.edu.cn}

\author{Liang Lin}
\affiliation{
\institution{Alibaba Cloud Computing}
}
\email{yibo.ll@alibaba-inc.com}

\author{Huan Li}
\authornote{Huan Li and Lidan Shou are the corresponding authors.}
\author{Gang Chen}
\affiliation{
\institution{Zhejiang University$^\dag$\authornote{The State Key Laboratory of Blockchain and Data Security}}
}
\email{{lihuan.cs, cg}@zju.edu.cn}

\begin{abstract}
Cloud-native serverless data warehouses achieve fine-grained elasticity by decoupling storage from compute, yet determining the optimal resource allocation for highly heterogeneous ad-hoc queries remains a formidable industrial challenge. Our analysis of production workloads in Alibaba AnalyticDB exposes a costly ``provisioning trap'': the fear of catastrophic resource depletion drives users to blindly over-provision resources, wasting immense monetary budgets without alleviating non-CPU bottlenecks (e.g., I/O saturation).
To break this impasse, we propose \ScaleSense, a proactive, query-level resource scaling framework. Specifically, it features a multi-faceted query encoder that jointly models plan topologies and hardware specifications. Crucially, a quantile-based resource predictor estimates multi-dimensional physical footprints, acting as a reliable safety net for optimal resource scaling. An auto-scaling controller then navigates the performance--cost Pareto frontier, dynamically tailoring allocations to specific business
priorities without requiring model retraining.
Evaluations on over 1.36 million production queries show that \ScaleSense achieves state-of-the-art prediction accuracy with good prediction interval coverage. By achieving a 76.7\% relative improvement in optimal resource configuration selection over the best baseline, this approach addresses the critical performance-cost trade-off while maintaining low-overhead inference latency, confirming its practical performance in production deployments. Under the performance-optimization policy, \ScaleSense satisfies user-defined performance requirements while reducing monetary cost by up to 5.22$\times$.
\end{abstract}

\maketitle

\pagestyle{\vldbpagestyle}
\begingroup\small\noindent\raggedright\textbf{PVLDB Reference Format:}\\
\vldbauthors. \vldbtitle. PVLDB, \vldbvolume(\vldbissue): \vldbpages, \vldbyear.\\
doi:\vldbdoi
\endgroup
\begingroup
\renewcommand\thefootnote{}\footnote{\noindent
This work is licensed under the Creative Commons BY-NC-ND 4.0 International License. Visit \url{https://creativecommons.org/licenses/by-nc-nd/4.0/} to view a copy of this license. For any use beyond those covered by this license, obtain permission by emailing \href{mailto:info@vldb.org}{info@vldb.org}. Copyright is held by the owner/author(s). Publication rights licensed to the VLDB Endowment. \\
\raggedright Proceedings of the VLDB Endowment, Vol. \vldbvolume, No. \vldbissue\%
ISSN 2150-8097. \\
\href{https://doi.org/\vldbdoi}{doi:\vldbdoi} \\
}\addtocounter{footnote}{-1}\endgroup

\ifdefempty{\vldbavailabilityurl}{}{
\vspace{.3cm}
\begingroup\small\noindent\raggedright\textbf{PVLDB Artifact Availability:}\\
The source code, data, and/or other artifacts have been made available at \url{\vldbavailabilityurl}.
\endgroup
}

\section{Introduction} \label{section:introduction}

\begin{figure}[ht]
	 \centering
  \includegraphics[width=0.45\textwidth, height=0.5\textheight]{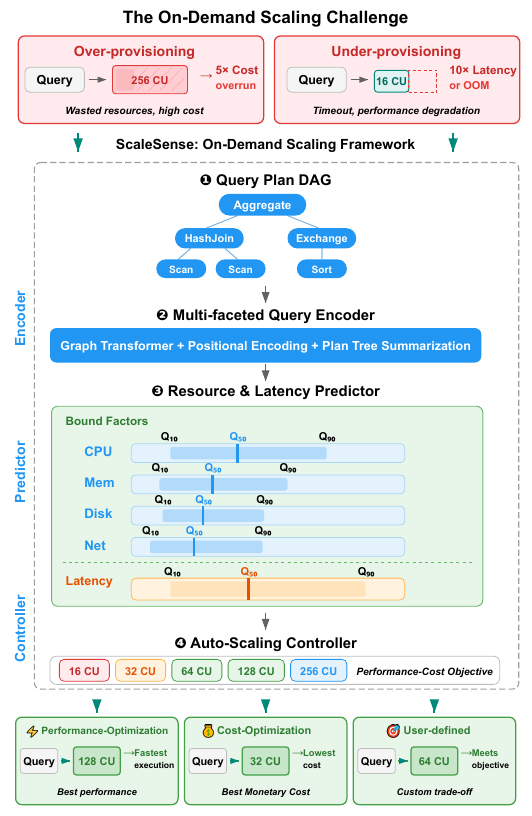}
	 \caption{The role of \ScaleSense.}
     \label{figure:ScaleSense_role}
\end{figure}

The emergence of cloud-native serverless architectures, exemplified by prominent systems such as Snowflake \cite{Snowflake}, Amazon Redshift \cite{Armenatzoglou2022Redshift}, and Alibaba Cloud AnalyticDB \cite{AnalyticDB}, has transformed data management by leveraging storage-compute disaggregation to achieve fine-grained elasticity. In these environments, fixed-size clusters are replaced by elastic compute pools that can scale dynamically to match fluctuating workloads. To simplify management, computational resources are increasingly abstracted into standardized units, such as AnalyticDB Compute Units (ACUs) \cite{AnalyticDBACU} or Redshift Processing Units (RPUs) \cite{RedshiftRPU}. This shift enables a transition from coarse-grained cluster resizing to granular, per-query resource allocation, offering more precise control over execution efficiency. However, this abstraction introduces a formidable processing challenge:
How can we predictively determine the \textbf{optimal resource configuration} for each ad-hoc query to strike a balance between execution performance and monetary cost?

Despite the flexibility of serverless billing, production environments are plagued by a profound mismatch between resource allocation and actual workload demands, forcing users into a costly \textbf{provisioning trap}. On one end of the spectrum, \emph{under-provisioning} leads to catastrophic Out-of-Memory (OOM) failures. Our previous study~\cite{Wu2026SafeLoad} revealed that over 3,000 queries suffer OOMs daily in a single production region. Although this is a statistically negligible fraction ($<0.0001\%$), each incident interrupts mission-critical workloads and squanders up to 17 CPU hours. In response to these reliability failures, users instinctively over-compensate by provisioning 2--3$\times$ the necessary resources as a crude safety margin, causing average cluster CPU utilization to plummet below 30\%, as reported by Google~\cite{autopilot2020} and Microsoft~\cite{resourcecentral2017}.
Crucially, this expensive \emph{over-provisioning} is often futile. Production traces demonstrate that query bottlenecks frequently shift between memory-intensive joins, I/O-bound scans, and network-constrained shuffles~\cite{dremel2020, fuxi2014}. \textbf{\textit{Consequently, blindly throwing more Compute Units (CUs) at an I/O-bound query merely burns monetary budget without delivering any latency reduction.}} Breaking this vicious cycle requires precise, multi-dimensional foresight at the per-query level.

Addressing these issues in the real production environment poses several critical technical challenges:
\begin{itemize}[leftmargin=*]
    \item \textbf{C1: Diverse Query Profiles.} Production workloads are characterized by an immense variety of ad-hoc queries with highly distinct and transient execution patterns, necessitating fine-grained, proactive prediction at the per-query level to accurately capture individual requirements before execution begins.

    \item \textbf{C2: Non-linear, Multi-dimensional Scaling Dynamics.} The relationship between allocated resources and query performance is likely non-linear. A query may be limited by different bottlenecks --- CPU, memory bandwidth, disk I/O, or network throughput --- at its different execution stages. Simply scaling up the CUs does not guarantee proportional performance gains due to Amdahl's Law and I/O saturation, complicating the search for an optimal performance-to-cost equilibrium.

    \item \textbf{C3: Multi-Objective Optimization under User Perception.} Resource provisioning in serverless environments is inherently subjective, as users often have conflicting preferences regarding performance and cost. Developing a universal policy is notoriously difficult, necessitating a provisioning framework that supports intuitive and tunable trade-off mechanisms, allowing users to explicitly define their tolerance for cost expansion and their requirements for performance acceleration.

    \item \textbf{C4: Reliability and Operational Constraints.} In production, a resource allocation framework must be both reliable and efficient. Models without confidence estimation are risky to deploy. At the same time, any inference overhead must be negligible to avoid delay in query execution.

\end{itemize}

Various approaches have been proposed for cloud resource management, yet they often fall short of addressing these four challenges simultaneously. Traditional reactive scaling methods \cite{CloudScalingSurvey} adjust capacity based on real-time telemetry, but suffer from an inherent detection lag that fails to prevent resource exhaustion.
Predictive scaling models \cite{Metis} forecast future demand from historical trends, but they struggle with the structural diversity of ad-hoc queries whose execution plans differ from historical patterns.
Some commercial platforms \cite{Snowflake, Armenatzoglou2022Redshift} utilize elastic auto-scaling and machine learning-driven workload management. While these systems can estimate query-level resource demands for admission control and memory allocation, their scaling logic remains primarily cluster-centric or based on coarse-grained resource tiers. They lack the ability to balance multi-dimensional footprints with user-tunable cost-performance trade-offs, which is necessary to optimize efficiency for ad-hoc workloads.
Optimization-based methods like CherryPick \cite{CherryPick} utilize Bayesian search to find ideal configurations, but their absolute reliance on multiple trial runs makes them impractical for production deployment.

To break this impasse, we propose \ScaleSense (\figurename~\ref{figure:ScaleSense_role}), an end-to-end, preference-aware auto-scaling framework that shifts the paradigm from deterministic, reactive provisioning to probabilistic, proactive query-level foresight.
First, to overcome workload heterogeneity (\textbf{C1}) and diverse hardware landscapes (\textbf{C2}), we design a Multi-faceted Query Encoder. Unlike traditional statistical models, it directly captures the hierarchical topology of execution plan Directed Acyclic Graphs (DAGs) and fuses them with runtime statistics and hardware specifications into a unified latent space.
Second, to handle non-linear bottlenecks (\textbf{C2}) and operational uncertainty (\textbf{C4}), we introduce a Quantile Resource Predictor. Moving beyond fragile point estimates, it explicitly predicts statistical bounds for four fundamental physical constraints. Equipped with a \emph{zero-classifier} for sparse data, it acts as a preventative feasibility filter, proactively pruning configurations prone to OOM and ensuring robust execution bounds for the remaining viable options.
Third, to resolve the multi-objective scaling problem under varying user preferences (\textbf{C3}), we introduce a preference-aware Auto-scaling Controller leveraging the Factor-Informed Hurwicz Criterion (FIHC) to formalize resource provisioning as a constrained multi-objective optimization problem. By incorporating both performance-oriented and cost-efficient operational policies, the controller identifies optimal resource specifications that adhere to user-defined performance and budget thresholds, enabling the selection of Pareto-optimal configurations tailored to diverse user priorities without retraining the prediction models.

Empirical evaluations on over 1.36 million production queries confirm \ScaleSense's effectiveness across all evaluation dimensions. It not only achieves state-of-the-art accuracy, reducing median Q-Error for resource estimation by up to 78\%, but fundamentally ensures operational safety by maintaining good prediction interval coverage.
In terms of auto-scaling efficiency, \ScaleSense achieves a 76.7\% improvement in CU configuration recommendation accuracy over the best baseline. Notably, under the performance-optimization policy, it satisfies user-defined performance requirements while reducing monetary cost by up to 5.22$\times$ compared to existing methods.

Our key contributions are summarized as follows:
\begin{itemize}[leftmargin=*]
    \item We design an on-demand auto-scaling framework that integrates a multi-faceted query encoder, a quantile-based resource predictor, and a preference-aware auto-scaling controller. \ScaleSense facilitates fine-grained performance--cost trade-offs, empowering users to dynamically navigate varying budget and latency constraints without the overhead of model retraining.

    \item Moving beyond black-box latency estimation, we develop a unified performance and resource prediction architecture that predicts query latency alongside four fundamental performance-bound factors. We uniquely integrate a zero-inflated classifier to handle extreme data sparsity, ensuring robustness across heterogeneous workloads.

    \item We pioneer a risk-controlled scaling mechanism based on Factor-Informed Hurwicz Criterion. By using calibrated prediction intervals as proactive safety guardrails, our framework predictively prunes configuration candidates that pose a high risk of memory pressure or hardware saturation.

    \item We conduct extensive evaluations on the TPC-DS benchmark and massive real-world production datasets. Results demonstrate that \ScaleSense achieves strong multi-dimensional prediction accuracy with negligible inference overhead.

\end{itemize}

The remainder of this paper is structured as follows: Section \ref{section:background} introduces the background. Section \ref{section:feature-analysis} distills empirical design insights, motivating the \ScaleSense framework detailed in Section \ref{section:solution}. Section \ref{section:evaluation} provides comprehensive evaluations, followed by related work in Section \ref{section:related_work}. Section~\ref{section:lessons} presents failure analysis and operational lessons. Section~\ref{section:conclusion} concludes this paper.
\section{Background} \label{section:background}
This section first describes the architectural design of serverless AnalyticDB with \ScaleSense integration, then formalizes the prediction and scaling problems addressed in this work.

\subsection{AnalyticDB's Serverless Query Execution}
\label{subsection:adb_overview}

\begin{figure}
	 \centering
  \includegraphics[width=0.47\textwidth]{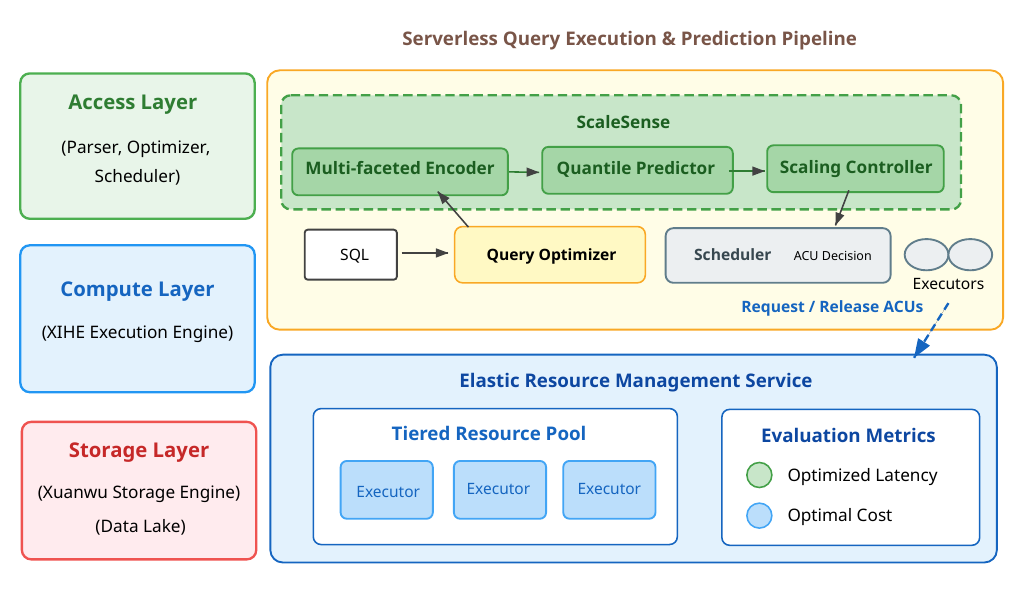}
	 \caption{The architecture of AnalyticDB, highlighting \ScaleSense as the proactive decision boundary between query optimization and resource scheduling.}
     \label{figure:fig_adb_architecture}
\end{figure}

AnalyticDB is a cloud-native, real-time data warehouse built to handle heterogeneous workloads with diverse resource demands.
To achieve high performance and elasticity, AnalyticDB follows a storage-compute disaggregated architecture, similar to industrial counterparts like Snowflake \cite{Snowflake}, and Amazon Redshift \cite{Armenatzoglou2022Redshift}.
As shown in Figure \ref{figure:fig_adb_architecture}, the system is organized into three functional layers. The \emph{Access Layer} handles SQL entry, connection management, and query optimization, employing a Parser and an Optimizer to transform SQL statements into physical execution plans.
The \emph{Compute Layer} executes query tasks using a distributed engine consisting of stateless executor nodes that can be scaled on demand, while the \emph{Storage Layer} consists of storage nodes from distributed storage systems to support high-throughput analytical scans.

To simplify cloud resource management, AnalyticDB abstracts computational capacity into \textbf{\emph{Compute Units}} (CUs).
Each CU represents a standardized bundle of CPU, memory, and I/O resources allocated at the per-query level.
AnalyticDB defines one CU as one CPU core paired with 4GB of main memory.
When a query is submitted, the system assigns a specific number of CUs for execution.
CU availability is ensured by the IaaS layer through an ECS warm pool that maintains pre-provisioned capacity for fast allocation.
This design provides second-level provisioning readiness, allowing pre-warmed resources to be assigned to a query almost instantaneously. Resource exhaustion and admission control are handled by the infrastructure scheduler, independent of the per-query CU sizing decision. Under extreme load, queries wait in the scheduler queue before CU allocation, and the sizing decision is invoked only after the query is admitted. Once allocated, each CU is provisioned as an isolated lightweight VM on ECS infrastructure, with CPU, memory, and I/O bandwidth reserved for the assigned query. Provisioning time is not billed to the customer, and after execution, the CU is returned to the warm pool for reuse in continuous workloads.

To achieve cost-intelligent scaling, \ScaleSense is injected directly into the critical execution path as a proactive decision boundary between the Access and Compute layers. When the Optimizer generates a physical DAG for a query, the \textbf{\emph{Multi-faceted Query Encoder}} captures its structural topology and statistical features. Rather than relying on reactive telemetry, the \textbf{\emph{Quantile Resource Predictor}} translates these features into probabilistic, multi-dimensional execution bounds. Then the \textbf{\emph{Auto-Scaling Controller}} evaluates user-defined preferences alongside the predicted resource distributions to dynamically determine the optimal CU configuration. Finally, this decision is passed to the Scheduler to provision the exact required capacity from the Tiered Resource Pool, while Executors carry out the actual execution.

\textbf{Deployment Status}.
The default performance optimization policy of the CU recommendation mechanism, which accelerates performance under a given monetary cost budget, is fully deployed in production to serve all customers, while its customizable recommendation feature --- balancing monetary cost and performance --- has been rigorously validated through controlled experiments and real customer workloads.

\begin{figure*}[ht]
	 \centering
  \includegraphics[width=1\textwidth]{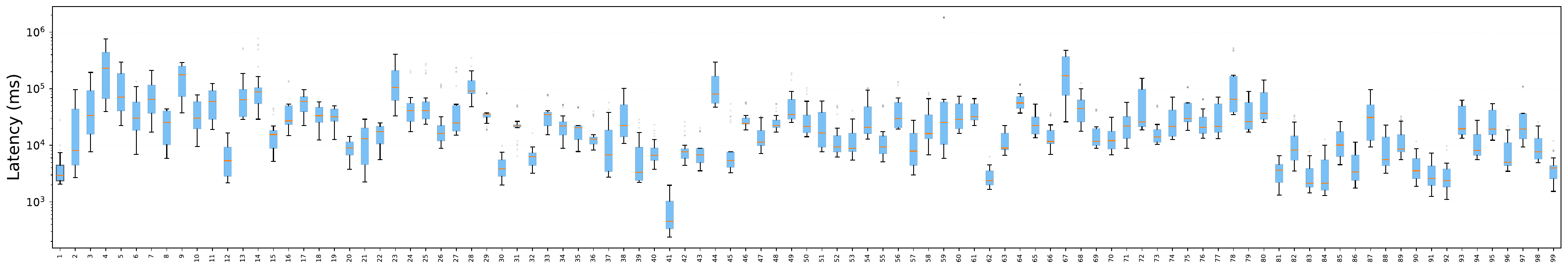}
	 \caption{Latency distribution of TPC-DS 1\,TB queries across six CU configurations (16, 32, 64, 128, 256, and 512 CUs).}
     \label{figure:tpcds_latency_variability}
\end{figure*}

\subsection{Problem Statement} \label{subsection:problem-statement}

Effective on-demand scaling requires precise query performance forecasting \emph{across diverse hardware configurations} ~\cite{zhang2023costintelligentdataanalyticscloud}. As Section~\ref{section:feature-analysis} demonstrates, ad-hoc queries rarely scale linearly due to shifting, query-specific hardware bottlenecks. To overcome the limitations of monolithic, black-box latency models, we decompose execution dynamics into observable dimensions. Guided by our empirical production analysis in Section~\ref{section:feature-analysis}, we explicitly isolate four physically grounded bottleneck indicators, namely (i)~CPU time, (ii)~peak memory size, (iii)~table scan volume (disk I/O), and (iv)~shuffle size (network I/O), to mechanistically forecast the ultimate optimization target: (v)~end-to-end query latency. We formulate the \emph{performance and resource estimation} problem as follows.

\begin{problem}[Performance and Resource Estimation]
\label{def:resource-estimation-problem}
Given a query $q$ and a candidate CU configuration $a \in \mathcal{A}$, the objective is to learn a predictive mapping $f_\theta(q, a) \rightarrow \hat{y}_d$ for each physical resource dimension $d \in \mathcal{D}_r = \{\textit{cpu},\, \textit{mem},\, \textit{disk},\, \textit{net}\}$, alongside the query latency $\hat{y}_{\textit{lat}}$. Crucially, these intermediate resource estimates $\{\hat{y}_d\}_{d \in \mathcal{D}_r}$ are not merely passive metrics; they serve as bottleneck-aware, foundational signals that explicitly guide the downstream latency prediction (detailed in Section~\ref{subsection:quantile-prediction}).
\end{problem}

For Problem~\ref{def:resource-estimation-problem},
our primary point-estimation metric is
\emph{Q-error}, defined as $\max(\hat{y}/y,\; y/\hat{y})$, which symmetrically penalizes over- and under-estimation of the predicted value.
We use Q-error only to evaluate the accuracy of the resource prediction.
The asymmetric operational consequences of prediction errors are handled downstream by the auto-scaling controller in Section~\ref{subsection:acu-selection}.

Equipped with these multi-dimensional forecasts, we must translate estimated latency and cost into actionable resource configurations from $\mathcal{A}$.
Instead of a fixed scoring function, we adopt a \emph{constraint-based optimization approach} that explicitly incorporates user-defined elasticity parameters to navigate the performance-cost trade-off. We formalize this \emph{on-demand scaling} problem as follows.

\begin{problem}[On-Demand Scaling with User Preference]
\label{def:user-defined-scaling-problem}
Let $\mathcal{A} = \{a_1, a_2, \ldots, a_k\}$ denote the discrete set of candidate CU configurations. For a query $q$ provisioned with configuration $a$, let $\textit{CU}(a)$ denote the assigned compute units, $L_a$ the predicted latency, and $C_a = \textit{CU}(a) \times L_a$ the projected monetary cost.

The objective is to select an optimal configuration $a^* \in \mathcal{A}$ that strictly adheres to the user's operational constraints. We characterize this trade-off using two user-defined parameters: the \textit{Cost Scaling Factor} ($\epsilon$) and the \textit{Performance Scaling Factor} ($\rho$), both normalized against a predefined baseline configuration $a_{base} \in \mathcal{A}$.
Depending on the operational priority, the system supports two complementary optimization policies:
\begin{enumerate}[leftmargin=*]
    \item \textbf{\emph{Performance Optimization (PO) Policy}:} The user seeks \textbf{performance speedup} within a specific budget expansion. The goal is to minimize execution latency while ensuring the cost remains within the tolerance $\epsilon$ and the speedup meets the target $\rho$:
    \begin{equation}
        a^* = \arg \min\nolimits_{a \in \mathcal{A}} \{ L_a \mid (C_a/C_{base} \le \epsilon) \wedge (L_{base}/L_a \ge \rho) \}.
    \end{equation}

    \item \textbf{\emph{Cost Optimization (CO) Policy}:} The user prioritizes \textbf{budget savings}, provided the performance degradation remains within a predefined acceptable threshold. The goal is to minimize the monetary cost while ensuring $\rho$ is met, and the cost ceiling $\epsilon$ is respected:
    \begin{equation}
        a^* = \arg \min\nolimits_{a \in \mathcal{A}} \{ C_a \mid (L_a/L_{base} \le \rho) \wedge (C_{base}/C_a \ge \epsilon) \}.
    \end{equation}
\end{enumerate}
\end{problem}

By unifying preferences into $\epsilon$ and $\rho$, our framework provides a consistent interface for users to navigate the Pareto front, regardless of whether their primary goal is acceleration or frugality.

Our monetary cost model follows the billing model of AnalyticDB Serverless~\cite{analyticdb_pricing}.
In this model, query execution consumes compute resources measured by CU-time, i.e., the allocated compute units multiplied by their usage time. Accordingly, $C_a$ captures the cost component of executing a query under CU configuration $a$, which is the component directly affected by \ScaleSense's per-query CU allocation decision.
This model is not intended to represent the entire cloud bill. In practice, users may also incur storage or data-transfer charges, and these factors depend on deployment-specific usage. Therefore, these factors are excluded from the optimization objective.
In our setting, CU resources are allocated from a warm pool, so provisioning latency is treated as an operational scheduling issue rather than a separately charged user cost.

\section{Empirical Design Insights} \label{section:feature-analysis}

\begin{figure*}[ht]
	 \centering
  \includegraphics[width=1\textwidth]{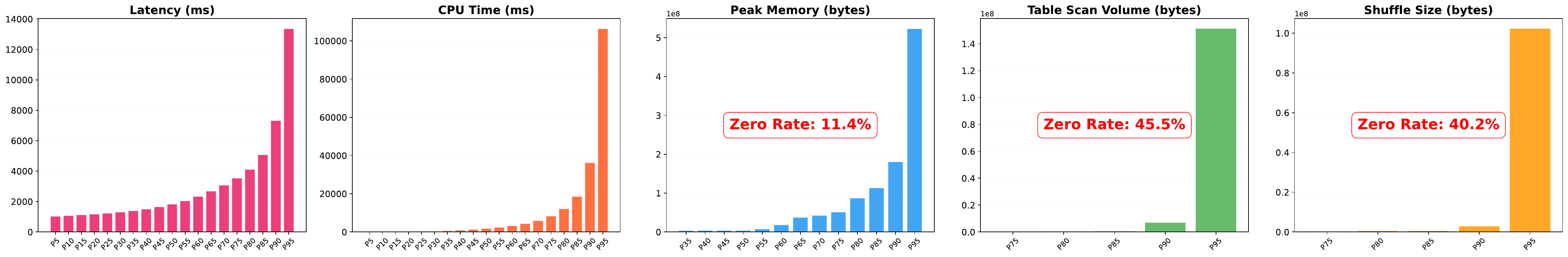}
	 \caption{Quantile distribution of query latency and key factors across the CPU, memory, disk I/O, and network dimensions.}
     \label{figure:distribution_bound_factors}
\end{figure*}

In this section, we derive the foundational architectural motivations for \ScaleSense through an empirical analysis of both the TPC-DS benchmark and massive production query logs.

As a classical OLAP benchmark, TPC-DS is well-suited to represent typical analytical query workloads. We execute all 99 TPC-DS queries at the 1\,TB scale factor under six CU configurations (16, 32, 64, 128, 256, and 512 CUs). Each query is run more than six times in strict isolation (with caches flushed and resources exclusively reserved), yielding the latency distributions shown in Figure~\ref{figure:tpcds_latency_variability}.

Across all queries, scaling from 16 to 512 CUs yields a latency reduction in 88\% of cases; however, scaling from 256 to 512 CUs alone yields a reduction in latency in only 53\% of cases. This saturation suggests that the \textbf{performance–resource scaling relationship is governed by distinct performance bounds across different CU ranges}.

\begin{table}[ht]
\centering
\footnotesize
\caption{Quantitative resource evidence across scaling tiers.}
\label{tab:scaling_mechanisms}
\footnotesize
\begin{tabular}{cc l rrr}
\toprule
\textbf{Bound Type} & \textbf{Query} & \textbf{Metric} & \textbf{16 CU} & \textbf{128 CU} & \textbf{512 CU} \\
\midrule

\multirow{4}{*}{CPU-Bound} & \multirow{4}{*}{\texttt{Q28}}
& \cellcolor{gray!20} CPU Time (Hours) & \cellcolor{gray!20} 2.83 & \cellcolor{gray!20} 15.53 & \cellcolor{gray!20} 30.00 \\
& & Peak Memory (GB) & 10.8 & 41.9 & 131.0 \\
& & Table Scan (GB) & 109.0 & 109.0 & 109.0 \\
& & Shuffle Size (GB) & 1.95 & 2.21 & 1.93 \\
\midrule

\multirow{4}{*}{Memory-Bound} & \multirow{4}{*}{\texttt{Q78}}
& CPU Time (Hours) & 5.17 & 12.69 & 14.97 \\
& & \cellcolor{gray!20} Peak Memory (GB) & \cellcolor{gray!20} 147.0 & \cellcolor{gray!20} 254.0 & \cellcolor{gray!20} 410.0 \\
& & Table Scan (GB) & 106.0 & 106.0 & 106.0 \\
& & Shuffle Size (GB) & 174.0 & 171.0 & 176.0 \\
\midrule

\multirow{4}{*}{Network-Bound} & \multirow{4}{*}{\texttt{Q23}}
& CPU Time (Hours) & 3.56 & 13.53 & 15.58 \\
& & Peak Memory (GB) & 35.2 & 90.0 & 280.0 \\
& & Table Scan (GB) & 211.0 & 211.0 & 211.0 \\
& & \cellcolor{gray!20} Shuffle Size (GB) & \cellcolor{gray!20} 370.0 & \cellcolor{gray!20} 368.0 & \cellcolor{gray!20} 364.0 \\
\midrule

\multirow{4}{*}{Disk I/O-Bound} & \multirow{4}{*}{\texttt{Q14}}
& CPU Time (Hours) & 11.47 & 14.56 & 17.08 \\
& & Peak Memory (GB) & 13.8 & 90.1 & 296.0 \\
& & \cellcolor{gray!20} Table Scan (GB) & \cellcolor{gray!20} 187.0 & \cellcolor{gray!20} 187.0 & \cellcolor{gray!20} 187.0 \\
& & Shuffle Size (GB) & 2.07 & 7.14 & 9.63 \\

\bottomrule
\end{tabular}
\end{table}

To isolate the specific mechanistic bounds dictating these non-linear scaling behaviors, we examine the resource footprints of the TPC-DS workload. Table~\ref{tab:scaling_mechanisms} presents quantitative evidence for four representative queries across scaling tiers, showing how different execution plans are gated by distinct physical constraints.
The \textit{CPU-bound} $\texttt{Q28}$ involves complex \texttt{COUNT(DISTINCT)} sub-queries where deduplication logic consumes extensive cycles, evidenced by CPU Time scaling from 2.83 to 30 hours at 512~CU while the shuffle size remains negligible (1.93~GB). In contrast, the \textit{memory-bound} $\texttt{Q78}$ utilizes multi-channel joins necessitating massive hash tables; its peak memory size (410~GB) is nearly 4$\times$ its 106~GB scan volume, making memory capacity critical to avoid disk spills. Moreover, $\texttt{Q23}$ is \textit{network-bound} due to global aggregations that trigger intense cross-node shuffling; its invariant 364~GB Shuffle Size significantly exceeds its 211~GB scan volume, establishing a physical latency floor that resists further compute expansion. Finally, $\texttt{Q14}$ is \textit{disk I/O-bound} and benefits from increased aggregate I/O bandwidth to handle its 187~GB scan volume, transitioning from an I/O-constrained bottleneck to a compute-balanced state.
These findings consolidate into our first core insight:
\begin{insight}\label{insight:tpc-ds}
Accurately predicting query latency requires accounting for two complementary sources of uncertainty: (i)~the {intrinsic run-to-run variability of query execution time} (Figure~\ref{figure:tpcds_latency_variability}), and (ii)~resource-configuration-dependent performance bounds along several primary dimensions --- \textbf{CPU, memory, disk I/O, and network} --- whose dominant factors are CPU time, peak memory size, table scan volume, and shuffle size, respectively (Table~\ref{tab:scaling_mechanisms}).
\end{insight}

Because latency exhibits an inherent run-to-run distributional range rather than collapsing to a single value, \textbf{a well-calibrated point estimator cannot adequately quantify execution uncertainty}.
To address this, \ScaleSense employs a quantile resource predictor (Section~\ref{subsection:quantile-prediction}) to explicitly capture the conditional latency distribution.
Since execution variability drops sharply outside extreme quantiles (e.g., the 10th and 90th percentiles), explicitly modeling those four dominant bottleneck factors substantially simplifies learning the complex resource–performance mapping.

To further understand these dynamics at scale, we conducted a quantile-level distributional analysis of query latency and the four key performance bounds, sampling one million long-running queries ($>1$ s) from the AnalyticDB production environment. The results, partitioned into 5\%-wide quantile bins, are presented in Figure~\ref{figure:distribution_bound_factors}.
As Figure~\ref{figure:distribution_bound_factors} illustrates, CPU time suffers from a tail effect even more pronounced than latency itself, driven by massive deviations in the upper quantiles. Conversely, peak memory size, table scan volume, and shuffle size exhibit distinct zero-inflation, with exact-zero values accounting for 11.4\%, 45.5\%, and 40.2\% of observations, respectively.
Notably, queries that record no disk I/O are mainly those whose storage reads are fully eliminated by filtering; for example, predicate filtering or runtime filters may prune all candidate data before scan materialization, so the final recorded table-scan bytes are zero.
We summarize the second core insight:
\begin{insight}\label{insight:bound-factor}
Performance-bound factors exhibit \textbf{extreme distributional skew} and \textbf{high sparsity}. While query latency and CPU time display severe heavy-tailed distributions, the I/O and memory dimensions contain massive exact-zero regions.
\end{insight}

This profound \textbf{distributional skew invalidates standard regression modeling}. First, to handle the high sparsity, the \ScaleSense resource predictor \emph{must} incorporate a zero-value classifier (Section~\ref{subsection:quantile-prediction}); otherwise, exact-zero targets will introduce extreme bias and distort the loss gradients for positive-valued samples. Second, because different resource bottlenecks trigger distinct adverse scenarios at their extremes (e.g., peak memory spikes cause fatal OOM crashes, whereas CPU spikes induce severe contention), our architecture must explicitly model the extreme quantile boundaries (e.g., $P_{10}$ and $P_{90}$) rather than just the median, providing the risk-aware safety net necessary for stable auto-scaling.

\begin{figure}[t!]
  \centering
  \includegraphics[width=\columnwidth]{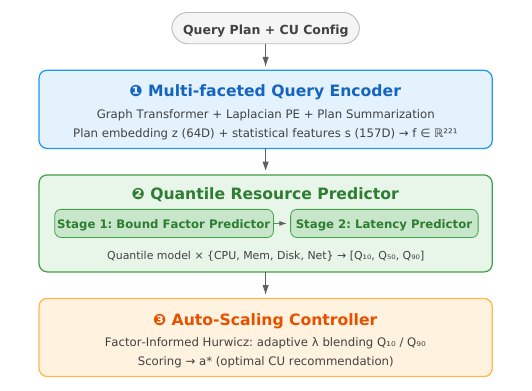}
  \caption{Overall framework of \ScaleSense.}
  \label{fig:ScaleSense-framework}
\end{figure}

\section{\ScaleSense Design} \label{section:solution}
This section presents the design of \ScaleSense, an on-demand scaling framework for serverless data warehouse environments. As illustrated in Figure~\ref{fig:ScaleSense-framework}, \ScaleSense operates as an integrated pipeline comprising three core components: (1) the Multi-faceted Query Encoder (Section~\ref{subsection:query-encoder}), which transforms raw query plans into compact vector representations; (2) the Quantile Resource Predictor (Section~\ref{subsection:quantile-prediction}), which provides multi-dimensional resource forecasts with calibrated uncertainty quantification; and (3) the Auto-Scaling Controller (Section~\ref{subsection:acu-selection}), which translates these probabilistic estimates into compute unit configuration decisions.

\subsection{Multi-faceted Query Encoder} \label{subsection:query-encoder}

\begin{figure}[t!]
  \centering
  \includegraphics[width=\columnwidth]{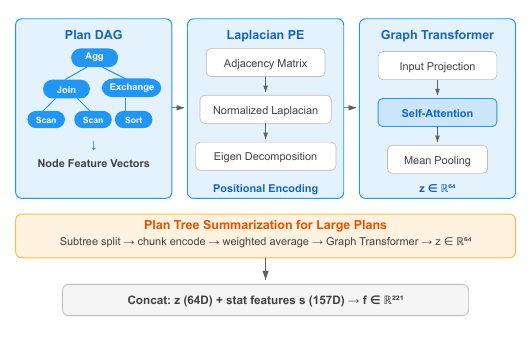}
  \caption{Architecture of the Multi-faceted Query Encoder.}
  \label{fig:plan-encoder}
\end{figure}

Query plans are naturally represented as directed acyclic graphs (DAGs), where nodes correspond to operators and edges represent data flow. Effectively encoding this structure is crucial for accurate prediction. As shown in Figure~\ref{fig:plan-encoder}, the encoder comprises three stages: node feature extraction with Laplacian Positional Encoding (LPE), a Plan Graph Transformer that produces a fixed-dimensional plan embedding, and a plan tree summarization strategy for handling large query plans.

\noindent\textbf{Node Feature Extraction with LPE}.
Graph neural networks have shown great promise in representation learning for structured data. Recently, Lyu et al.~\cite{Lyu2025Graph} demonstrated that \emph{Laplacian Positional Encoding} (LPE) significantly improves the ability of Transformers to capture global topological properties of graphs. Query plans share similar structural characteristics with the graphs studied in \cite{Lyu2025Graph}---they are directed acyclic graphs where the relative position of a node (e.g., whether a Join is at the bottom or top of the tree) fundamentally alters its resource impact.

Building on this insight, we \emph{adapt} the LPE-based Transformer architecture for query plan encoding. Unlike prior work that uses tree convolutions~\cite{Marcus2021Bao} (local-only) or height encodings~\cite{Zhao2022QueryFormer} (vertical-only), our approach leverages the graph Laplacian to capture the full global topology of the query plan.

For each operator node (denoted by index $v$) in the plan DAG, we extract an 11-dimensional feature vector:
\begin{equation}
    \mathbf{x}_v = [\text{op\_type}, \log(\text{rows}), \log(\text{bytes}), \text{flags}, \text{structural}],
\end{equation}
encompassing (1) \emph{$\text{op\_type}$}: Operator type index, normalized by the total number of operator categories (19 types including HashJoin, TableScan, Aggregate, Exchange, Sort, etc.); (2) \emph{$\log(\text{rows}), \log(\text{bytes})$}: Log-transformed estimated output rows and estimated output bytes from optimizer statistics, each normalized by the respective maximum observed value; (3) \emph{$\text{flags}$}: Binary indicators for operator categories (e.g., is\_join) and (4) \emph{$\text{structural}$}: Tree depth, parallelism degree, and selectivity estimates.

To capture the global structure of the plan DAG, we compute LPE based on the graph Laplacian matrix. While spectral analysis typically treats the graph as undirected, this does not result in critical information loss for our task. The global topology (e.g., depth, width, bottleneck nodes) captured by the Laplacian spectrum is direction-independent but crucial for resource estimation. Furthermore, the Transformer's self-attention mechanism implicitly learns directional dependencies during training, complementing the structural bias provided by LPE~\cite{Lyu2025Graph}.

Given the adjacency matrix $\mathbf{A} \in \mathbb{R}^{n \times n}$ of the plan DAG (treated as undirected for spectral analysis), we compute the normalized Laplacian:
$
    \tilde{\mathbf{L}} = \mathbf{I} - \mathbf{D}^{-1/2}\mathbf{A}\mathbf{D}^{-1/2},
$
where $\mathbf{D}$ is the degree matrix. We compute the eigen-decomposition $\tilde{\mathbf{L}} = \mathbf{U}\mathbf{\Lambda}\mathbf{U}^\top$ and use the eigenvectors corresponding to the $k$ smallest non-zero eigenvalues as positional encodings
$
    \mathbf{PE}(v) = [\mathbf{U}_{v,1}, \mathbf{U}_{v,2}, \ldots, \mathbf{U}_{v,k}].
$

The eigenvectors of the Laplacian encode structural information about node centrality, connectivity patterns, and graph partitioning. Nodes with similar structural roles (e.g., leaves, roots, bottleneck operators) receive similar positional encodings, enabling the model to generalize across query plans with different sizes but similar topological patterns.

\noindent\textbf{Plan Graph Transformer Architecture}.
The Plan Graph Transformer processes the sequence of node features augmented with positional encodings (Figure~\ref{fig:plan-encoder}), i.e.,
$
    \mathbf{h}_v^{(0)} = \mathbf{W}_x \mathbf{x}_v + \mathbf{W}_p \mathbf{PE}(v).
$
Multiple Transformer encoder layers apply self-attention across all nodes:
$
    \mathbf{h}^{(\ell+1)} = \text{TransformerLayer}(\mathbf{h}^{(\ell)}).
$

The final plan embedding is obtained via mean pooling over all node representations, i.e.,
$
    \mathbf{z} = \frac{1}{|V|} \sum\nolimits_{v \in V} \mathbf{h}_v^{(L)}.
$
This aggregation strategy produces a fixed-dimensional plan embedding $\mathbf{z} \in \mathbb{R}^{64}$ regardless of the number of operators in the plan. Mean pooling provides consistent performance across varying plan sizes while being computationally efficient.

\noindent\textbf{Plan Tree Summarization for Large Plans}.
Production query plans can contain hundreds of operators, making direct Transformer encoding prohibitively expensive ($\mathcal{O}(N^2)$ self-attention). To handle such plans efficiently, we introduce a \emph{plan tree summarization} strategy that partitions the plan DAG into manageable chunks while preserving local structural information.

\begin{algorithm}[t]
\DontPrintSemicolon
\footnotesize
\caption{Encoding with Plan Tree Summarization}\label{alg:plan-summarization}
\KwIn{Plan DAG nodes $V$, chunk size budget $C$}
\KwOut{Plan embedding $\mathbf{z} \in \mathbb{R}^{64}$}
\If{$|V| \leq C$}{
    \Return $\text{PlanGraphTransformer}(V)$\;
}
Compute subtree size $\text{sz}(v)$ for each $v \in V$ via bottom-up traversal\;
$\mathcal{S} \leftarrow \emptyset$; $\text{buf} \leftarrow \emptyset$\;
\ForEach{node $v$ in DFS order}{
    \eIf{$\text{sz}(v) \leq C$}{
        \If{$|\text{buf}| + \text{sz}(v) > C$ \textbf{and} $\text{buf} \neq \emptyset$}{
            $\mathcal{S} \leftarrow \mathcal{S} \cup \{\text{buf}\}$; $\text{buf} \leftarrow \emptyset$\;
        }
        $\text{buf} \leftarrow \text{buf} \cup \text{Subtree}(v)$\;
    }{
        Add $v$ alone to $\text{buf}$; push children to stack\;
    }
}
$\mathcal{S} \leftarrow \mathcal{S} \cup \{\text{buf}\}$\;
\ForEach{chunk $c_i \in \mathcal{S}$}{
    Build local adjacency $\mathbf{A}_i$ (intra-chunk edges only)\;
    $\mathbf{z}_i \leftarrow \text{PlanGraphTransformer}(c_i, \mathbf{A}_i)$\;
}
$\mathbf{z} \leftarrow \sum_i \frac{|c_i|}{\sum_j |c_j|} \cdot \mathbf{z}_i$ \;
\Return $\mathbf{z}$\;
\end{algorithm}

The key idea is to split the plan along \emph{subtree boundaries}: we perform a DFS traversal, compute the sub-tree size of each node, and greedily assign complete sub-trees to chunks until a size budget $C$ (default 200 nodes) is reached. This ensures that each chunk retains the parent-child relationships within its subgraph, preserving the local topology that the Transformer relies on. Each chunk is then independently encoded by the Plan Graph Transformer, and the final plan embedding is obtained by a weighted average of chunk embeddings, where the weight of each chunk is proportional to its node count. Algorithm~\ref{alg:plan-summarization} details this procedure.

\noindent\textbf{Complexity Analysis}. This approach reduces the per-query encoding cost from $\mathcal{O}(N^3)$ to $\mathcal{O}(NC^2)$, where $N$ is the total number of operators and $C$ is the chunk budget. The dominant cost in full-plan encoding is the Laplacian eigen-decomposition $\mathcal{O}(N^3)$, followed by Transformer self-attention ($\mathcal{O}(N^2 d)$). With plan tree summarization, each of the $M = \lceil N/C \rceil$ chunks independently computes its local Laplacian PE ($\mathcal{O}(C^3)$ each) and runs self-attention ($\mathcal{O}(C^2 d)$ each), yielding a total cost of $\mathcal{O}(NC^2 + NCd)$. Since $C$ is a fixed constant (default 200), the complexity is \emph{linear} in $N$ for fixed $C$.

After obtaining the structural feature vector of the query plan via the Graph Transformer, we concatenate it with a plan-level statistics feature vector (adopted from prior work~\cite{Wu2026SafeLoad}) and
hardware configuration information (e.g., the CU configuration) to form the final representation. A detailed description of the plan statistics features is provided elsewhere~\cite{SafeBench}.

\subsection{Quantile Resource Predictor} \label{subsection:quantile-prediction}

\begin{figure}[ht]
  \centering
  \includegraphics[width=\columnwidth]{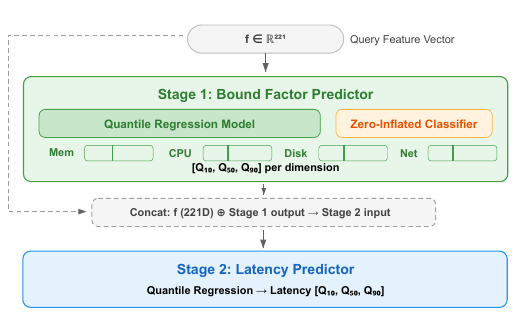}
  \caption{Architecture of the Quantile Resource Predictor.}
  \label{fig:bound-model}
\end{figure}

The core insight of \ScaleSense is decomposing resource prediction into two stages, as illustrated in Figure~\ref{fig:bound-model}: performance-bound factor prediction (first stage) and latency prediction (second stage).

\noindent\textbf{Bound Factor Predictor (Stage 1)}.
The Bound Factor Predictor predicts the performance bounds of a query across four dimensions, with uncertainty quantification for all dimensions. Critically, predictions are conditioned on the target CU configuration, as some bounds (notably Disk I/O) depend on available memory. The input is $\mathbf{f} = [\mathbf{z} \| \mathbf{s}]$, where $\mathbf{z}$ is the 64-dimensional plan embedding and $\mathbf{s}$ is a 157-dimensional vector of statistical features extracted from the optimizer plan (including operator counts, cardinality estimates, hardware configuration, and aggregate plan statistics). The concatenated input $\mathbf{f} \in \mathbb{R}^{221}$ captures both structural and statistical information.

We employ \textbf{\emph{gradient-boosted quantile regression}}~\cite{Romano2019quantile} to predict three quantiles for \emph{each} dimension (Figure~\ref{fig:bound-model}):
\begin{equation}
    [Q_{10}^{(d)}, Q_{50}^{(d)}, Q_{90}^{(d)}] = [\text{XGB}_{\tau}^{(d)}(\mathbf{f})]_{\tau \in \{0.1, 0.5, 0.9\}},
\end{equation}
where $\forall d \in \{\text{cpu}, \text{mem}, \text{disk}, \text{net}\}$.

For each dimension and quantile level, an independent XGBoost (XGB) model is trained on log-transformed targets. Monotonicity ($Q_{10} \leq Q_{50} \leq Q_{90}$) is enforced at inference time by clipping:
\begin{equation}
    Q_{10}^{(d)} \leftarrow \min(Q_{10}^{(d)}, Q_{50}^{(d)}), \quad Q_{90}^{(d)} \leftarrow \max(Q_{90}^{(d)}, Q_{50}^{(d)}).
\end{equation}

We adopt the 10th and 90th percentiles to construct an 80\% prediction interval. This choice is informed by our empirical analysis (Section~\ref{section:feature-analysis}), which shows that the bulk of execution variability concentrates within this range, while the tails beyond Q10 and Q90 may be caused by transient runtime noise.
This design is critical for risk-controlled allocation: memory under-estimation leads to OOM failures, which are far more catastrophic than CPU under-estimation (which merely causes slowdowns). By providing prediction intervals for memory usage, \ScaleSense enables the scheduler to provision memory conservatively for memory-sensitive queries.

To ensure model fidelity, all \textbf{\emph{training labels}} are empirically acquired directly from query execution logs, avoiding error-prone heuristic decomposition or estimation. The multi-dimensional physical constraints are defined as follows: (i) $B_\text{cpu}$, the total CPU time (ms) captured via execution telemetry; (ii) $B_\text{mem}$, the peak memory footprint (bytes) recorded by task-level instrumentation; (iii) $B_\text{disk}$, the disk I/O volume (bytes) aggregated from table scan counters; and (iv) $B_\text{net}$, the network I/O payload extracted from shuffle counters.

Unlike approaches that decompose observed latency into latent resource components, we predict known, measurable quantities recorded by the query engine during execution. This makes the system debuggable: if a prediction is wrong, we can identify which dimension was mispredicted.

\noindent\textbf{Latency Predictor (Stage 2)}.
The Latency Predictor is a quantile regression model that maps the Stage~1 outputs and features to end-to-end latency \emph{with uncertainty quantification}. It takes as input the concatenation of scaled features and the full Stage~1 quantile predictions, and outputs latency quantiles directly:
\begin{equation}
    [\hat{L}^{Q_{10}}, \hat{L}^{Q_{50}}, \hat{L}^{Q_{90}}] = \text{LatencyPredictor}([\mathbf{f}_{scaled} \| \text{flatten}(\hat{\mathbf{Q}}^{S1})]),
\end{equation}
where $\hat{\mathbf{Q}}^{S1} \in \mathbb{R}^{4 \times 3}$ denotes the Stage~1 quantile predictions across four dimensions and three quantile levels.

This design has two key advantages: 1) End-to-end Supervision: The model is trained directly on observed latency, avoiding the pitfalls of intermediate efficiency coefficients that can lead to uninterpretable compensation effects; 2) Native Uncertainty: The Stage~2 model produces its own calibrated prediction intervals for latency, rather than relying on propagation of Stage~1 quantiles through a deterministic mapping.

Both stages use \textbf{\emph{quantile loss}} (pinball loss) as the training objective. For a target quantile level $\tau$:
\begin{equation}
    \mathcal{L}_\tau(y, \hat{y}) = \begin{cases}
        \tau (y - \hat{y}) & \text{if } y \geq \hat{y} \\
        (1-\tau) (\hat{y} - y) & \text{if } y < \hat{y}
    \end{cases}.
\end{equation}

XGBoost natively optimizes the pinball loss through second-order gradient boosting. For multiple quantiles $\tau \in \{0.1, 0.5, 0.9\}$, separate models are trained for each quantile level.

The 80\% prediction interval is constructed as $[Q_{10}, Q_{90}]$. The median $Q_{50}$ serves as the point estimate. The interval width $Q_{90} - Q_{10}$ quantifies prediction uncertainty: wider intervals indicate higher uncertainty, suggesting more conservative resource allocation.

For resource dimensions with high zero rates---peak memory (${\sim}11.4\%$ zeros), table scan volume (${\sim}45.5\%$ zeros), and shuffle size (${\sim}40.2\%$ zeros)---we employ a \emph{Hurdle Model} approach. A lightweight gradient-boosted classifier first predicts whether the resource consumption is zero. For samples classified as zero, the corresponding Stage~1 quantile predictions are set to zero before being passed to Stage~2. The quantile regression model is trained only on non-zero samples for these dimensions (via a loss mask that excludes zero-valued targets), allowing it to focus on predicting the conditional distribution given non-zero consumption.

\subsection{Auto-Scaling Controller} \label{subsection:acu-selection}

Given the quantile predictions from the quantile resource predictor for each candidate CU configuration, the Auto-Scaling Controller selects the configuration that minimizes the optimization objective defined in Problem~\ref{def:user-defined-scaling-problem}. The key challenge is: which latency estimate should be used as $L_a$ in the CU configuration selection? Using the median $\hat{L}^{Q_{50}}$ ignores prediction uncertainty, while using the upper bound $\hat{L}^{Q_{90}}$ is overly conservative. We propose a \emph{Factor-Informed Hurwicz} criterion that adaptively blends optimistic and pessimistic latency estimates based on resource pressure signals from Stage~1.

\noindent\textbf{Candidate Enumeration.}
For each query $q$, we evaluate candidate CU configurations $\mathcal{A} = \{a_1, a_2, \ldots, a_k\}$. For each candidate $a$, the Two-Stage Quantile Resource Predictor produces latency quantiles $(\hat{L}^{Q_{10}}_a, \hat{L}^{Q_{50}}_a, \hat{L}^{Q_{90}}_a)$ and Stage~1 factor quantiles $(\hat{W}^{Q_{10}}_{a,d}, \hat{W}^{Q_{50}}_{a,d}, \hat{W}^{Q_{90}}_{a,d})$ for each resource dimension $d$.

\noindent\textbf{Factor-Informed Hurwicz Criterion.}
The classical Hurwicz criterion~\cite{Hurwicz1951, Trummer2021SkinnerDB,Helmer2025Hurwicz} blends the best-case and worst-case outcomes using a fixed optimism parameter. We extend this idea by making the blending parameter \emph{adaptive} --- informed by both the user preference $\alpha$ and the resource pressure signals from Stage~1 predictions.

For each candidate configuration $a$, we compute a per-query blending parameter $\lambda(q, a) \in [0, 1]$ that controls the interpolation between the optimistic ($Q_{10}$) and pessimistic ($Q_{90}$) latency estimates:
\begin{equation}
    \hat{L}_a = \lambda(q, a) \cdot \hat{L}^{Q_{10}}_a + (1 - \lambda(q, a)) \cdot \hat{L}^{Q_{90}}_a
    \label{eq:hurwicz-latency}.
\end{equation}

The blending parameter is composed of two signals:
\begin{equation}
\lambda(q, a) = \alpha \cdot \lambda_{\text{pressure}}(q, a)
  + (1 - \alpha) \cdot \lambda_{\text{position}}(q, a),
    \label{eq:lambda-blend}
\end{equation}
where $\alpha$ is the user preference coefficient from Problem~\ref{def:user-defined-scaling-problem}.

\noindent\textbf{Resource Pressure Signal} ($\lambda_{\text{pressure}}$).
This signal captures how heavily the candidate configuration is loaded relative to other candidates. We compute per-CU resource pressure for memory and CPU using the pessimistic (Q90) Stage~1 predictions:
\begin{equation}
    p_{\text{mem}}(a) = \frac{\hat{W}^{Q_{90}}_{a,\text{mem}}}{\text{CU}(a)}, \quad
    p_{\text{cpu}}(a) = \frac{\hat{W}^{Q_{90}}_{a,\text{cpu}}}{\text{CU}(a)}.
\end{equation}
Each pressure value is min-max normalized across candidates, and the pressure signal is:
\begin{equation}
    \lambda_{\text{pressure}}(q, a) = 1 - \tfrac{1}{2}\!\left(\tilde{p}_{\text{mem}}(a) + \tilde{p}_{\text{cpu}}(a)\right),
\end{equation}
where $\tilde{p}$ denotes the normalized pressure. Configurations with lower per-CU resource pressure receive higher $\lambda_{\text{pressure}}$ (more optimistic), reflecting the intuition that under-loaded configurations are less likely to experience performance degradation. We adopt CPU and memory for the Resource Pressure Signal because these two dimensions are directly tied to the CU configuration in AnalyticDB, whereas the other two factors (disk I/O and network I/O) are not governed by the CU specification.

\noindent\textbf{Factor Position Signal} ($\lambda_{\text{position}}$).
This signal captures the skewness of the Stage~1 prediction intervals. For each resource dimension $d$, we compute the relative position of the median within its prediction interval:
\begin{equation}
\text{rp}_d(a) = \left[\frac{\hat{W}^{Q_{50}}_{a,d} - \hat{W}^{Q_{10}}_{a,d}}
                      {\hat{W}^{Q_{90}}_{a,d} - \hat{W}^{Q_{10}}_{a,d}}\right]_0^1.
\end{equation}
The position signal is the complement of the average relative position:
\begin{equation}
    \lambda_{\text{position}}(q, a) = 1 - \frac{1}{|\mathcal{D}_r|} \sum_{d \in \mathcal{D}_r} \text{rp}_d(a).
\end{equation}
When the median is close to $Q_{10}$ (left-skewed interval), $\text{rp}_d$ is small and $\lambda_{\text{position}}$ is high, indicating that the model is confident the true value is near the optimistic end. When the median is close to $Q_{90}$, $\text{rp}_d$ is large and $\lambda_{\text{position}}$ is low, signaling higher downside risk.

Under loose constraints ($\alpha = 0$), the dominant risk is prediction error; $\lambda_{\text{position}}$ steers $\hat{L}_a$ toward $Q_{90}$ when prediction intervals are right-skewed, guarding against over-optimism. Under tight constraints ($\alpha = 1$), the dominant risk is
resource contention; $\lambda_{\text{pressure}}$ penalizes configurations with high per-CU load, preventing tolerance violations from memory
spilling or CPU throttling.

\section{Evaluation} \label{section:evaluation}
We evaluate \ScaleSense across prediction accuracy, auto-scaling efficiency, and system overheads, addressing three core questions:
\begin{itemize}[leftmargin=7.8mm]
\item[{\bf EQ1}] \uline{Prediction Performance Comparison (Section \ref{subsection:accuracy})}:
How accurately does \ScaleSense predict query latency and resource consumption compared to state-of-the-art baselines?

\item[{\bf EQ2}] \uline{Efficiency of Auto-Scaling (Section \ref{subsection:auto-scaling})}:
How effectively does \ScaleSense achieve user-preferred performance-cost trade-offs compared to the baselines?

\item[{\bf EQ3}] \uline{Time Overheads (Section \ref{subsection:overheads})}:
What are the empirical training and inference time costs of \ScaleSense?

\end{itemize}

\subsection{Experimental Settings} \label{subsection:settings}

\noindent\textbf{Hardware Setup.}
All queries are executed on Alibaba Cloud AnalyticDB instances. All model training and inference experiments are conducted on a single machine equipped with an Intel Xeon Platinum 8163 CPU (64 cores, 2.50\,GHz), and 252\,GB RAM.

\noindent\textbf{Baselines.}
Our experimental evaluation is organized into two parts. For query performance and resource estimation (Part I), we compare against four competitive baselines: GTN, XGB, MLP, and LGBM. For on-demand scaling evaluation (Part II), we also include a heuristic rule (LSR) and SS-naive.
We introduce each baseline as follows:
\begin{itemize}[leftmargin=*]
    \item \textbf{Graph Transformer Network (GTN)~\cite{Lyu2025Graph, Lyu2024GTN}}: Recent studies~\cite{Lyu2025Graph, Lyu2024GTN} have shown that Graph Transformer Networks achieve state-of-the-art performance in query plan encoding and are widely adopted for cardinality estimation and latency prediction.

    \item \textbf{XGBoost-based Predictor (XGB)~\cite{RAIS2024, Wu2026SafeLoad}}: XGBoost is extensively used for latency and memory demand estimation in RAIS~\cite{RAIS2024}, the state-of-the-art intelligent scaling system. A recent study on memory-overloading queries~\cite{Wu2026SafeLoad} further demonstrates that XGBoost is the best-performing model for memory estimation. This baseline applies XGBoost regression on the query feature vector, following the same details as SafeLoad~\cite{Wu2026SafeLoad}.

    \item \textbf{MLP-based Predictor (MLP)~\cite{Quader2024LearnedWMP}}: Multi-layer perceptrons are widely used for resource estimation due to their low online inference overhead. LearnedWMP~\cite{Quader2024LearnedWMP} employs an MLP for memory demand prediction. This baseline applies an MLP on the query feature vector.

    \item \textbf{LightGBM-based Predictor (LGBM)}: LightGBM is commonly regarded as an alternative to XGBoost and frequently serves as a baseline in the resource estimation literature~\cite{Quader2024LearnedWMP, Wu2026SafeLoad}.

    \item \textbf{Linear Scaling Rule (LSR)}: LSR is a heuristic strategy that selects the minimum CU configuration under cost-optimization policy, and the maximum CU under performance-optimization policy. Despite its simplicity, this rule is effective for queries whose performance-resource scales approximately linearly.

    \item \textbf{\ScaleSense-naive (SS-naive)}: An ablated variant of \ScaleSense that removes the Factor-Informed Hurwicz Criterion, relying solely on the median (Q50) point estimate for CU selection. This baseline isolates the contribution of the uncertainty-aware decision module.

\end{itemize}

\noindent\textbf{Benchmark.}
The evaluation of \ScaleSense considers both representative production workloads and the open TPC-DS 1TB benchmark. For query performance and resource estimation experiments, we evaluate on two real-world production datasets, large-scale dataset D1 and mid-scale dataset D2, both comprising complex queries with execution times exceeding 1 second, making them well-suited for estimation and on-demand scaling tasks.
D1 contains 1 million real SQL queries executed across 117 distinct CU configurations.
D2 is constrained to six hardware configurations (16, 32, 64, 128, 256, and 512 CUs) and contains 360K real SQL queries.

For the performance and resource estimation experiments, D1 and D2 are each split into training and test sets at an 8:2 ratio for in-distribution evaluation. Since real-world on-demand scaling scenarios typically involve complex, out-of-distribution (OOD) queries, we adopt the open TPC-DS 1TB benchmark as the test set for auto-scaling evaluation. Specifically, we restrict the candidate CU configurations to the same six settings (16, 32, 64, 128, 256, and 512 CUs), train all methods on D2, and evaluate their CU recommendation effectiveness on TPC-DS 1TB.

\noindent\textbf{Evaluation Metrics.}
Q-error is widely regarded as the standard metric~\cite{Rieger2025T3}, defined as $\max(\hat{y}/y,\; y/\hat{y})$. We report the median and 90th-percentile Q-errors for all predicted dimensions. Unless otherwise specified, \ScaleSense uses Q50 and real label values to calculate evaluation metrics.
Q-error measures prediction accuracy only. Memory under-estimation may cause OOM, while over-estimation mainly increases cost, and quantiles with FIHC handle this asymmetry. Relative Q-error is $\frac{\text{Q-error of method } M}{\text{Q-error of best baseline}}$, so values below 1 indicate improvement.
Since \ScaleSense produces quantile intervals, we additionally report two interval-based metrics: \emph{Coverage Rate (CR)}, the fraction of true values falling within the predicted interval, and \emph{Interval Width (IW)}, the mean width of the prediction intervals.
Additionally, we evaluate the zero-inflated classifiers using classification accuracy on zero-valued samples.
To evaluate the effectiveness of on-demand scaling solutions, we adopt \emph{Constraint Satisfaction Accuracy (CSA)} as the primary metric.
CSA measures the fraction of feasible queries for which the recommended CU configuration satisfies both the performance constraint and the cost constraint simultaneously.

\noindent\textbf{Implementation Details.}
We reuse the full set of plan-level features introduced in SafeLoad~\cite{Wu2026SafeLoad}. \ScaleSense incorporates a Plan Graph Transformer that uses 8 Laplacian eigenvectors, 2 Transformer layers with a hidden dimension of 64, and 4 attention heads.
For the plan tree summarization procedure, we set the chunk size to 200 operator nodes, a threshold empirically determined from AnalyticDB production statistics. Both the Bound Factor Predictor (Stage~1) and the Latency Predictor (Stage~2) employ XGBoost-based quantile regression with a maximum tree depth of 10, a learning rate of 0.03, and subsample and column sample ratios of 0.8.
The Graph Transformer and XGBoost settings are adopted from prior work~\cite{Lyu2025Graph, Wu2026SafeLoad}.
Although XGBoost-based quantile regressors extrapolate conservatively beyond the training range, \ScaleSense mitigates this limitation through log-scale targets, broad production coverage, and a fallback to the maximum CU when prediction intervals become wide.

\begin{figure*}[t!]
	 \centering
  \includegraphics[width=1\textwidth, height=0.3\textheight]{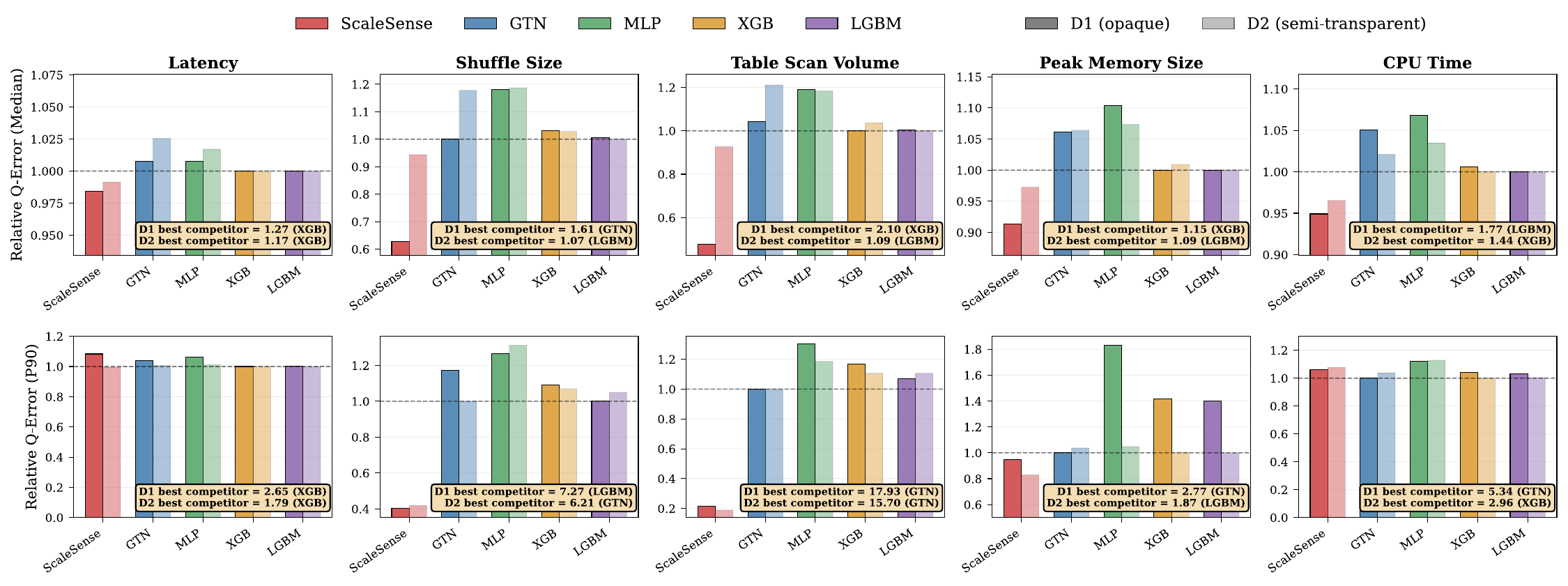}
	 \caption{Relative Q-Error of methods compared to the best baseline (lower is better) on the real production datasets D1 and D2.} \label{figure:qerror_comparison_merged}
\end{figure*}

\subsection{Prediction Performance Comparison (EQ1)} \label{subsection:accuracy}
We comprehensively evaluate \ScaleSense against all baselines on five prediction targets --- query latency, CPU time, peak memory size, table scan volume, and shuffle size --- using the real production datasets D1 and D2. In addition, we report the performance of the quantile interval prediction unique to \ScaleSense.

\noindent\textbf{Prediction Accuracy.}
As shown in Figure~\ref{figure:qerror_comparison_merged}, on the large-scale production dataset D1, \ScaleSense achieves the best Q-Error median across all prediction targets. For Q-Error P90, which reflects tail-case prediction quality, \ScaleSense attains the best results on table scan volume, shuffle size, and peak memory size.
Specifically, for table scan volume, \ScaleSense reduces the Q-Error median by 52.4\% and the Q-Error P90 by 78.3\% relative to the best baseline, demonstrating a substantial prediction advantage.
For shuffle size, \ScaleSense reduces the Q-Error median by 37.3\% and the Q-Error P90 by 59.8\% relative to the best baseline, again showing a significant improvement.
For peak memory size, \ScaleSense improves the Q-Error median by 8.7\% over the best baseline, with a marginal increase in Q-Error P90.
For the time-related metrics, latency and CPU time, \ScaleSense maintains the best Q-Error median with modest improvements of 1.6\% and 5.1\% over the best baseline, respectively, while exhibiting a slight degradation in Q-Error P90.
Overall, \ScaleSense achieves the best prediction performance for the majority of queries across all prediction targets on the production dataset D1.

As shown in Figure~\ref{figure:qerror_comparison_merged}, on the mid-scale production dataset D2, the reduced diversity of hardware configurations compared to D1 narrows the variation in extreme cases at the P90 quantile, making the advantage of \ScaleSense more pronounced. \ScaleSense achieves the best Q-Error median and P90 on every prediction target except for the Q-Error P90 of CPU time.
Specifically, for table scan volume, \ScaleSense reduces the Q-Error median by 7.3\% and the Q-Error P90 by 81.1\% relative to the best baseline.
For shuffle size, \ScaleSense improves the Q-Error median by 5.6\% and the Q-Error P90 by 58.1\% over the best baseline.
For peak memory size, \ScaleSense improves the Q-Error median by 2.8\% over the best baseline, while also achieving a 17.1\% improvement in Q-Error P90.

\noindent\textbf{Efficiency of Quantile Prediction.}
As shown in Table~\ref{table:coverage_zero_classifier}, \ScaleSense not only provides accurate point estimates but also predicts Q10--Q90 quantile intervals as prediction bounds. Across all five dimensions on both D1 and D2, the coverage rates range from 68\% to 86\% with narrow interval widths, confirming that the predicted intervals reliably capture the true values. These bounds enable the Factor-Informed Hurwicz
criterion to quantify resource uncertainty and make more robust CU selection decisions, which is the key advantage over point-estimate-only baselines.

\noindent\textbf{Zero-Inflated Classification.}
\ScaleSense also achieves good accuracy in zero-inflated classification. For peak memory size, the classification accuracy reaches 0.94 on D1 and 0.81 on D2. For table scan volume, it attains 0.93 on D1 and 0.96 on D2. For shuffle size, the accuracy is 0.97 on D1 and 0.85 on D2.

\begin{table}[t!]
        \centering
        \caption{Quantile prediction and zero-inflated classification performance of \ScaleSense on D1 and D2.}
        \resizebox{\columnwidth}{!}{
        \begin{tabular}{@{}l|lccclll@{}}
\toprule
Dataset & Target & Q-error (Median) & Q-error P90 & Coverage Rate & Acc. & Avg. IW &  \\ \midrule
\multirow{5}{*}{D1} & Latency & 1.25 & 2.87 & 78.87\% & \textbackslash{} & 5.37 s &  \\
 & Peak Memory Size & 1.05 & 2.62 & 68.03\% & 0.94 & 0.15 GB &  \\
 & Table Scan Volume & 1.00 & 3.89 & 83.03\% & 0.93 & 0.39 GB &  \\
 & Shuffle Size & 1.01 & 2.92 & 85.70\% & 0.97 & 0.21 GB &  \\
 & CPU Time & 1.68 & 5.67 & 78.84\% & \textbackslash{} & 69.95 s &  \\
\midrule
\multirow{5}{*}{D2} & Latency & 1.16 & 1.78 & 78.13\% & \textbackslash{} & 2.05 s &  \\
 & Peak Memory Size & 1.06 & 1.55 & 78.22\% & 0.81 & 0.06 GB &  \\
 & Table Scan Volume & 1.01 & 2.97 & 82.85\% & 0.96 & 0.15 GB &  \\
 & Shuffle Size & 1.01 & 2.60 & 82.25\% & 0.85 & 0.18 GB &  \\
 & CPU Time & 1.39 & 3.19 & 77.02\% & \textbackslash{} & 31.03 s &  \\
\bottomrule
\end{tabular}
        }
        \label{table:coverage_zero_classifier}
\end{table}

\noindent\textbf{Summary.}
\ScaleSense consistently achieves the best or near-best Q-Error median across all five prediction targets on both D1 and D2, with particularly pronounced improvements on I/O-related dimensions such as table scan volume and shuffle size. Beyond point estimation, its quantile interval predictions achieve good coverage with narrow interval widths, confirming that \ScaleSense provides accurate, uncertainty-aware resource predictions suitable for downstream scaling decisions.

\subsection{Efficiency of Auto-Scaling (EQ2)} \label{subsection:auto-scaling}
We evaluate all scaling solutions on the TPC-DS 1\,TB benchmark under the constraint-based formulation defined in Problem~\ref{def:user-defined-scaling-problem}. For each of the 99 TPC-DS queries, each method must recommend a target CU configuration that satisfies the user-specified constraints on both the performance scaling factor $\rho$ and the cost scaling factor $\epsilon$.
We give six constraint settings that cover two complementary policies. For \textbf{Performance Optimization (PO)}, the baseline is 16\,CUs (the minimum configuration) and the candidate targets are \{32, 64, 128, 256, 512\}\,CUs; the user demands a speedup $\rho \in [3.5, 4.0]\times$ while tolerating a cost increase $\epsilon \in [2.5, 3.0]\times$.  For \textbf{Cost Optimization (CO)}, the baseline is 128\,CUs (a mid-range configuration) and the candidates are \{16, 32, 64, 256, 512\}\,CUs; the user demands a cost saving $\epsilon \in [1.5, 2.0]\times$ while tolerating a latency degradation $\rho \in [1.1, 1.3]\times$.

These ranges are chosen to reflect practical production scenarios: PO settings target compute-intensive queries that benefit from substantial scale-up, while CO settings target over-provisioned queries where moderate scale-down yields significant savings with minimal performance impact.
Under the PO policy, the sum of the user-expected scaling ratios exceeds the heuristic threshold of 5, so \ScaleSense assigns $\alpha = 0$; under the CO policy, the user expectations are more conservative, and \ScaleSense accordingly assigns $\alpha = 1$.
Within each policy, we progressively tighten the constraints across three settings to stress-test each method's robustness.

\begin{figure}[t!]
	 \centering
  \includegraphics[width=0.47\textwidth]{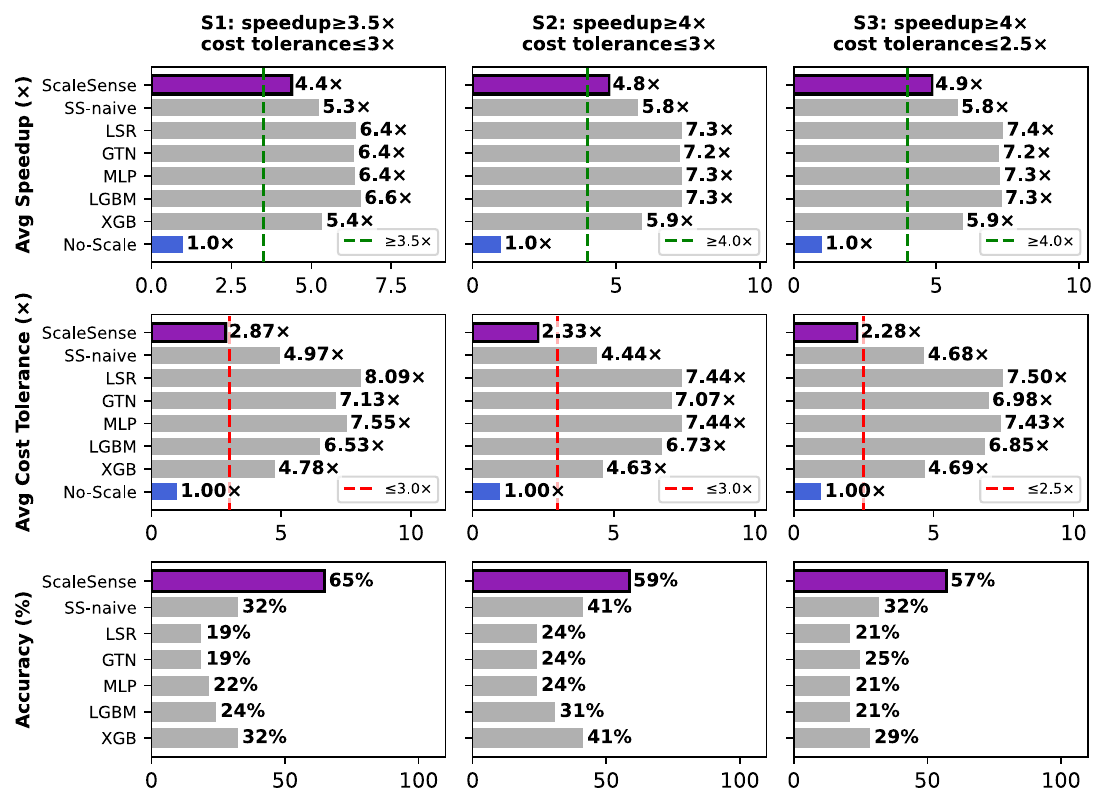}
	 \caption{Performance optimization (base CUs = 16\,CUs). } \label{figure:performance_optimization}
\end{figure}

\noindent\textbf{Performance Optimization (PO).}
Figure~\ref{figure:performance_optimization} reports the constraint satisfaction accuracy and the average speedup and cost ratio achieved by each method under three PO settings (baseline = 16\,CUs, targets $\in$ \{32, 64, 128, 256, 512\}). The green and red dashed lines indicate the speedup and cost tolerance thresholds, respectively. For example, the setting ($\rho \geq 3.5\times$, $\epsilon \leq 3\times$) requires at least a $3.5\times$ speedup while the cost must not exceed $3\times$ the baseline. Across all three settings, \ScaleSense achieves the highest accuracy, satisfying the user constraints on 57.1\%--64.9\% of feasible queries, compared to 28.6\%--41.4\% for the best baseline (XGB). On average, \ScaleSense improves over the best baseline by 26.1 percentage points (pp), corresponding to a relative improvement of 80.6\%. Crucially, \ScaleSense maintains an average cost ratio within the user-specified tolerance (e.g., $\epsilon = 2.87\times$ under a $3\times$ budget), whereas all baselines exceed the cost constraint on average (e.g., XGB averages $4.78\times$), indicating that they tend to over-provision resources. All baselines achieve higher total speedups ($3.6$--$4.4\times$) than \ScaleSense ($2.7\times$), but at the expense of a $4.7$--$7.2\times$ cost increase, far exceeding the user budget. In contrast, \ScaleSense keeps the total cost increase at $3.4$--$3.6\times$, demonstrating that it selects moderate configurations that balance speedup and cost rather than blindly maximizing performance.

\begin{figure}[t!]
	 \centering
  \includegraphics[width=0.47\textwidth]{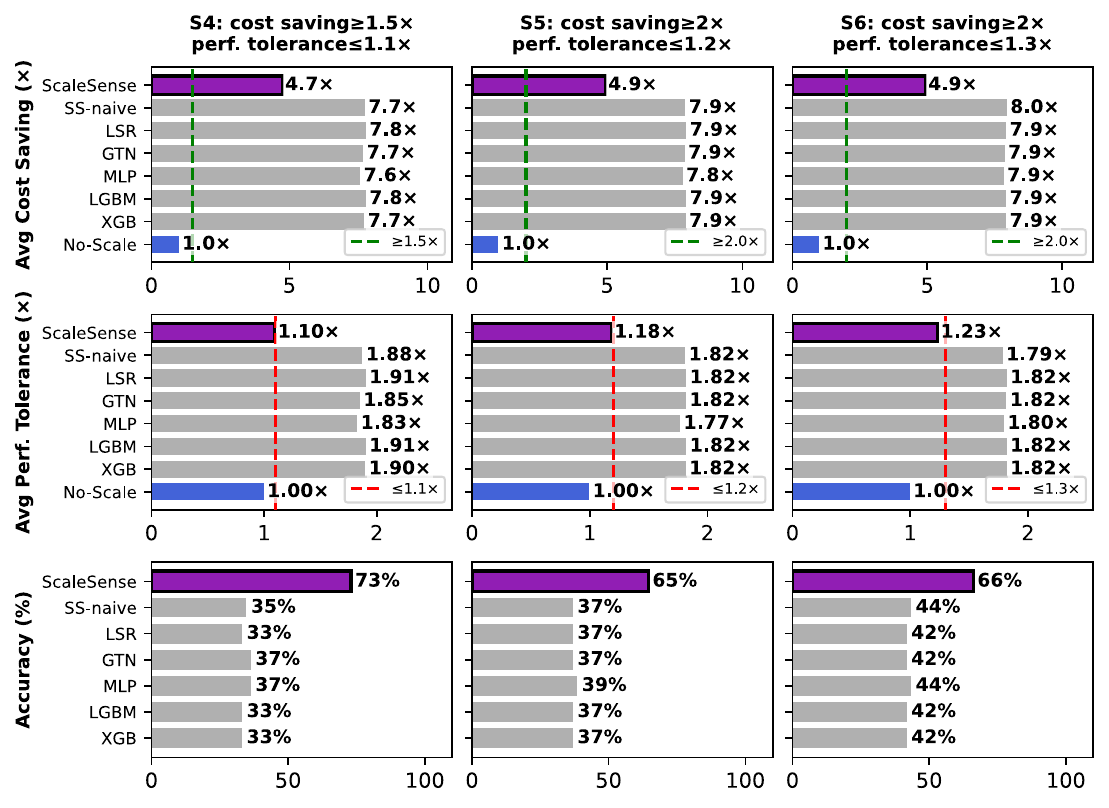}
	 \caption{Cost optimization (base CUs = 128\,CUs).} \label{figure:cost_optimization}
\end{figure}

\noindent\textbf{Cost Optimization (CO).}
Figure~\ref{figure:cost_optimization} reports the results under three CO settings (baseline = 128\,CUs, targets $\in$ \{16, 32, 64\} CUs). For example, the setting ($\epsilon \geq 1.5\times$, $\rho \leq 1.1\times$) requires at least a $1.5\times$ cost saving while the latency degradation must not exceed $1.1\times$. \ScaleSense achieves 64.5\%--73.0\% accuracy, compared to 36.5\%--43.5\% for the best baseline (MLP), an average improvement of 28.3\,pp. Notably, \ScaleSense keeps the average latency degradation close to the tolerance bound (e.g., $\rho = 1.09\times$ under a $1.1\times$ constraint), while all baselines incur substantially higher slowdowns (e.g., MLP averages $1.83\times$), reflecting their inability to identify configurations that save cost without excessive performance loss.
All baselines achieve a $3.1\times$ total cost saving but incur a $2.6\times$ total slowdown, indicating that they uniformly select the lowest CU (16) regardless of the constraint. \ScaleSense achieves a more moderate $1.8\times$ cost saving with only $1.8$--$1.9\times$ slowdown.
This is precisely why the saving target is not the main discriminative factor in
Figure~\ref{figure:cost_optimization}:
exceeding the target is easy if a method always chooses the smallest configuration, but doing so usually violates the latency constraint.
The key question is whether a method can realize non-trivial savings \emph{without} breaking the user-specified SLA, which is where \ScaleSense differs from the baselines.

\noindent\textbf{Impact of the Factor-Informed Hurwicz Criterion.}
To isolate the contribution of the uncertainty-aware FIHC module, we compare \ScaleSense with SS-naive, an ablated variant that uses only the median (Q50) point estimate for CU selection. Under the six settings (with CO candidates restricted to \{16, 32, 64\}), the Factor-Informed Hurwicz criterion improves CSA by an average of 27.13\,pp over SS-naive. The gain is +24.89\,pp in performance-optimization settings and +29.37\,pp in cost-optimization settings. FIHC also outperforms two simpler static robust alternatives: relative to always using Q10 and Q90, it improves CSA by +27.71\,pp and +17.59\,pp on average, respectively. These results show that the controller gain does not come from choosing a single conservative or optimistic quantile; it comes from factor-informed adaptation between quantiles according to policy semantics and resource pressure.

\noindent\textbf{Summary.}
Across all six constraint settings, \ScaleSense outperforms the best baseline by an average of 27.2\,pp in constraint satisfaction accuracy (relative improvement of 76.7\%). The Factor-Informed Hurwicz criterion accounts for the majority of this gain (+27.1\,pp over SS-naive), confirming that FIHC is the key enabler for effective on-demand scaling. Notably, under the performance-optimization policy, it satisfies user-defined performance requirements while reducing monetary cost by up to 5.22$\times$ compared to existing methods.

\subsection{Time Overheads (EQ3)} \label{subsection:overheads}
We analyze the training and inference overheads of \ScaleSense and all baselines.

\noindent\textbf{Training Overheads.}
On the D1 and D2 training sets comprising 800K and 288K queries, respectively, \ScaleSense completes training in 294\,s and 160\,s. This is slower than XGB (26\,s and 16\,s) and LGBM (112\,s and 85\,s), but faster than MLP (585\,s and 196\,s) and GTN (630\,s and 202\,s). The training cost of \ScaleSense falls between the tree-based and neural network baselines, as it employs XGBoost as the underlying model for quantile interval prediction. Nevertheless, these offline training costs are modest and remain entirely acceptable even for daily retraining scenarios.

\noindent\textbf{Inference Overheads.}
On the D1 and D2 test sets containing 200K and 72K queries, respectively, \ScaleSense completes inference in approximately 4\,s and 2\,s. All baselines, being lightweight models, also finish within 1--4\,s. In all cases, the inference overhead amounts to less than 1\% of the average query execution time and is therefore negligible in practice.

\section{Related Work} \label{section:related_work}

\noindent\textbf{Intelligent Auto-Scaling.}
Existing cloud data warehouses employ various workload management and scaling strategies.
Snowflake~\cite{Snowflake} leverages independent virtual warehouses for workload isolation and elastic capacity.
Microsoft's Moneyball~\cite{Poppe2022Moneyball} proactively predicts per-database pause/resume patterns  to reduce resume latency after idle periods, trading off quality of service and operating cost at the database level. Regarding predictive scaling in analytical warehouses, AWS Redshift RAIS~\cite{RAIS2024, Redshift-WLM} uses proactive, predictive algorithms that learn workload patterns and query complexity to adjust resources before degradation occurs. At a coarser provisioning granularity, Doppler~\cite{Cahoon2022Doppler} recommends right-sized Azure SQL PaaS targets for workload migration using low-level resource statistics and price-performance ranking, while Lorentz~\cite{Glaze2024Lorentz} recommends SKUs for newly provisioned services from customer profile data and continuous feedback when workload traces are unavailable.
In contrast, \ScaleSense introduces predictive CU resizing methodologies through multidimensional resource estimation.
Furthermore, \ScaleSense operates at per-query granularity with uncertainty quantification, enabling finer-grained resource allocation decisions.

\noindent\textbf{Query Resource Estimation.}
Early work used kernel canonical correlation analysis for memory prediction~\cite{Ganapathi2009MemoryPrediction}. Recent ML-based approaches include XGBoost on SQL text features~\cite{Tang2021Twitter} and CASA's statistical feature extraction~\cite{Zeyl2024CASA}. LinkedIn's QPP study~\cite{song2025evaluating} evaluated TLSTM and TCNN on industrial OLAP workloads, extending to CPU time prediction. These methods primarily produce point estimates for single metrics. \ScaleSense advances the state-of-the-art by: (1) predicting key performance bound factors (CPU, memory, disk, network), (2) providing calibrated prediction intervals via quantile regression to enable transferable performance bound estimation across diverse hardware configurations.

\noindent\textbf{Learned Query Optimization.}
Learned query optimization has emerged as a promising direction for improving database performance.
QueryFormer~\cite{Zhao2022QueryFormer} and DACE~\cite{liang2024dace} augment the Transformer architecture using height encoding and tree-structured attention masks, respectively. Lero~\cite{Zhu2023Lero} and LEON~\cite{LEON2023} employ pairwise query plan ranking rather than exact cost estimation for robust query optimization. A comparative study~\cite{Cong2024} systematically analyzed plan encoding techniques across these methods.
Learned cost models (LCMs) in this line of work focus on estimating the abstract cost of candidate query plans, whereas \ScaleSense predicts hardware-dependent resource consumption and latency across candidate CU configurations.

\section{Lessons and Future Directions}
\label{section:lessons}
This section summarizes failure analysis and operational lessons learned from building \ScaleSense in an industrial serverless data warehouse, AnalyticDB.

\noindent\textbf{Lesson 1: Deployment-oriented analysis is essential when full replay is infeasible.}
In production, privacy constraints prevent access to the underlying customer data. As a result, we cannot replay the same user query under multiple CU configurations to obtain ground-truth labels for CU selection. Controlled benchmarks such as TPC-DS are therefore indispensable for end-to-end evaluation under comparable conditions. At the same time, deployment-oriented analysis of real customer workloads remains necessary to verify that the learned recommendations translate into practical value.

\noindent\textbf{Lesson 2: Failure analysis is useful for constructing an effective feedback mechanism.}
Under-provisioning CUs for complex queries may cause resource exhaustion and execution failure. In our failure case study, TPC-DS 1TB query Q67 fails at 16 CUs because its large-scale shuffle and global sort operators create substantial memory pressure. This illustrates why CU selection should explicitly model resource pressure, rather than relying on latency point estimates alone, in modules such as the Factor-Informed Hurwicz Criterion. More importantly, the system should include a dedicated failure-aware fallback mechanism.

\noindent\textbf{Lesson 3: Robust relative ranking matters more than absolute accuracy.}
Out-of-distribution robustness is more important than in-distribution accuracy: preserving the relative ranking across candidate CU configurations, together with uncertainty-aware fallback, is sufficient to make safe and effective scaling decisions.

\section{Conclusion}
\label{section:conclusion}
In this paper, we present \ScaleSense, an on-demand scaling framework for cloud-native serverless data warehouses that addresses the challenge of selecting appropriate CU configurations. By jointly predicting query latency and multi-dimensional resource consumption with quantile uncertainty estimates, \ScaleSense explicitly models the performance--cost trade-off and translates it into actionable
scaling decisions aligned with user-defined preferences. Experimental evaluation demonstrates that \ScaleSense achieves state-of-the-art constraint-satisfaction accuracy in CU recommendation.

\begin{acks}
This research is supported by the NSFC under Grants No. U24A201401 and No. 62402420, and also supported by CCF-Aliyun Apsara Research Fund (No.~CCF-Aliyun2024008). We thank the anonymous reviewers and the AnalyticDB team for their valuable feedback and suggestions.
\end{acks}

\bibliographystyle{ACM-Reference-Format}
\bibliography{main/references}

\end{document}